# Dimming of the Mid- 20th Century Sun

Peter Foukal[1]

**Advances in understanding of the bright white light (WL) faculae measured at the Royal Greenwich Observatory (RGO) from 1874-1976 suggest that they offer a more direct measure of solar brightening by small – diameter photospheric magnetic flux tubes than do chromospheric proxies. Proxies such as the area of Ca K plages, the Mg II index, or the microwave flux include many dark photospheric structures as well as pores and sunspots . Our reconstruction of total solar irradiance (TSI) based on the WL faculae indicates that the sun dimmed by almost 0.1% in the mid- 20th century rather than brightening significantly as represented in previous reconstructions. This dimming at the sun's highest activity level since the 17th century is consistent with the photometric behavior observed in somewhat younger sun – like stars. The prolonged TSI decrease may have contributed more to the cooling of climate between the 1940's and 1960's than present models indicate**.

Our information on total solar irradiance (TSI) variation prior to the beginning of space-bornee measurements in 1978 is based mainly on reconstructions of the dimming by dark sunspots and brightening by photospheric faculae using a regression technique[1,2]. The technique requires that the contribution to TSI of the facular broad – band radiation be proportional to their emission from the overlying chromospheric layer. These chromospheric brightenings are called plages; they are more easily observed than the photospheric faculae, which are visible only near the solar limb (Fig.1).

Time series of the plage area variation have recently become available through digitization of daily solar images in the Ca II K line made beginning in the early 20 th century [3]. Shorter time series exist of e.g. the 10.7 cm microwave flux or Lyman alpha irradiance. The longest time series used to estimate this bright contribution is the sunspot number, R, which extends back to the early 17th century. Its value is dominated by the hundreds of tiny spots present at times of high solar activity. Although these are dark in photospheric radiation, they have bright chromospheres, so R tracks the chromospheric irradiances well and is also used as a proxy of the bright contribution[4].

Photometric imaging in the infrared [5,6] and at higher angular resolution in the visible [7,8] has shown that only the smallest plage elements are bright in photospheric radiation, at all positions on the disk. Much of the bright chromospheric emission, including that from larger plages (magnetic flux $> 2 \times 10^{18}$ Mx) actually marks flux tubes that are *dark* in photospheric radiations except when viewed near the limb. As illustrated in Fig.1, the chromospheric radiations are bright even over most sunspots, and spot areas are included in plage indices [9]. This progression of decreasing photospheric brightness with increasing flux tube cross section is expected from their dynamics[10]. While all WL faculae mark plages, not all plages mark WL faculae, so use of these proxies significantly over- estimates the bright contribution to TSI.

This contribution to TSI can be measured more directly by observing the white light (WL) faculae seen near the solar limb in Fig.1. These are the actual bright photospheric structures that produce the TSI contribution. Their daily area variation was measured by the Royal Greenwich Observatory (RGO) over the 102- year period between 1874 – 1976.[11]

[1]Heliophysics, Inc., Nahant, Massachusetts , 01908, USA.

The values, $A_f$, that we use here are the projected areas expressed as fractions of the photospheric disk. The technique used for their measurement and the error sources have been discussed elsewhere[12, 14]. The visibility of the WL faculae only near the limb presents no disadvantage if our interest is in the annual mean TSI which averages over solar rotations and is of main relevance in climate studies.

The main obstacle to their use has been the puzzling time behavior of their areas compared to the sunspot number or the chromospheric indices [10]. Their annual mean areas are positively correlated with the Ca K plage area, $A_{pn}$, at low to medium levels of solar activity. This can be seen in Fig.2 over the solar activity range between 1916 -1976. But the surprising finding is the *inverse* relation between the amplitudes of the three largest cycles 17, 18 and 19 in WL facular areas and in Apn. (The regular daily Ca K images from Mt Wilson Observatory from which the plage area index, $A_{pn}$, was derived first became available in late 1915).

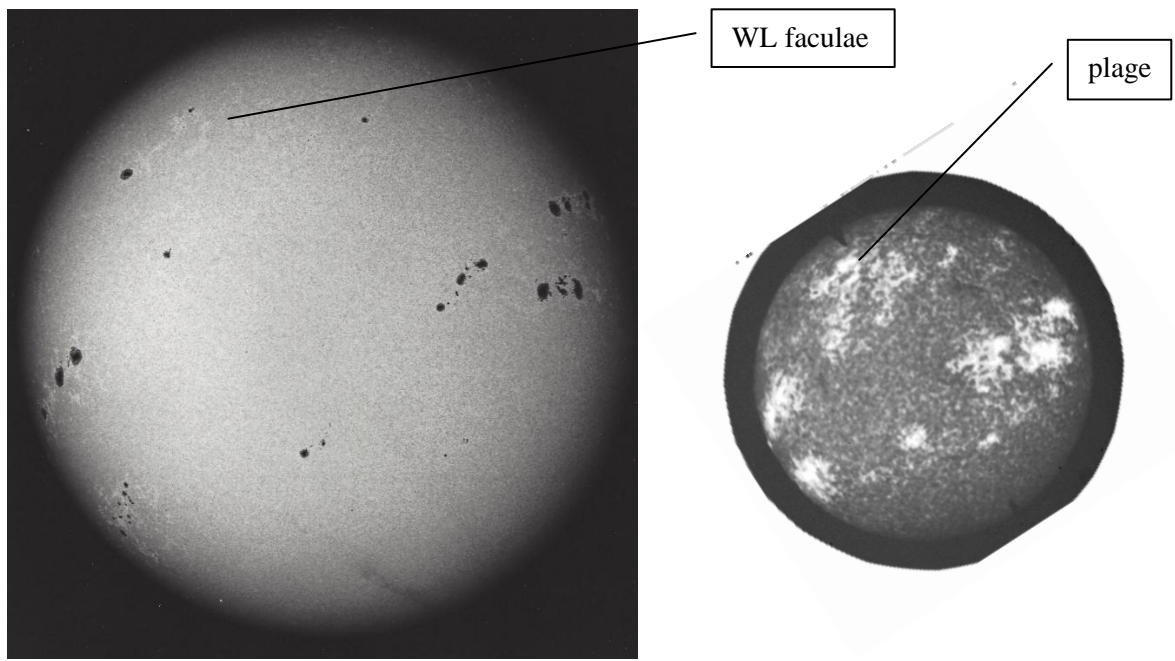

Fig.1 The solar disk imaged in photospheric white light (left) and in chromospheric Ca K radiation (right). WL faculae are seen best near the left limb; Ca K plages cover them and also most of the sun spots. Images obtained at Mt Wilson Observatory on Sept.15, 1957 near the peak of cycle 19.

This change in sign of the correlation was difficult to understand while the WL faculae were assumed to be simply the photospheric segment of a vertical magnetic flux tube whose upper atmosphere was observed as a chromospheric plage. We now know from better photometry that the WL faculae and the chromospheric indices measure overlapping but not identical classes of structures.

We also now have a better physical understanding of why the correlation changes sign. Study of the area ratio, $A_f/A_s$, of WL faculae and spots has shown that it remains constant over most of the sun's activity range, but decreases at the highest activity levels encountered in cycles 18 and 19 [12-14]. So the solar dynamo seems to favor larger – diameter flux tubes at the highest solar activity levels. Such behavior is consistent with observations of very large spots on younger, more active solar-type stars [15] and with recent dynamo calculations.[16]

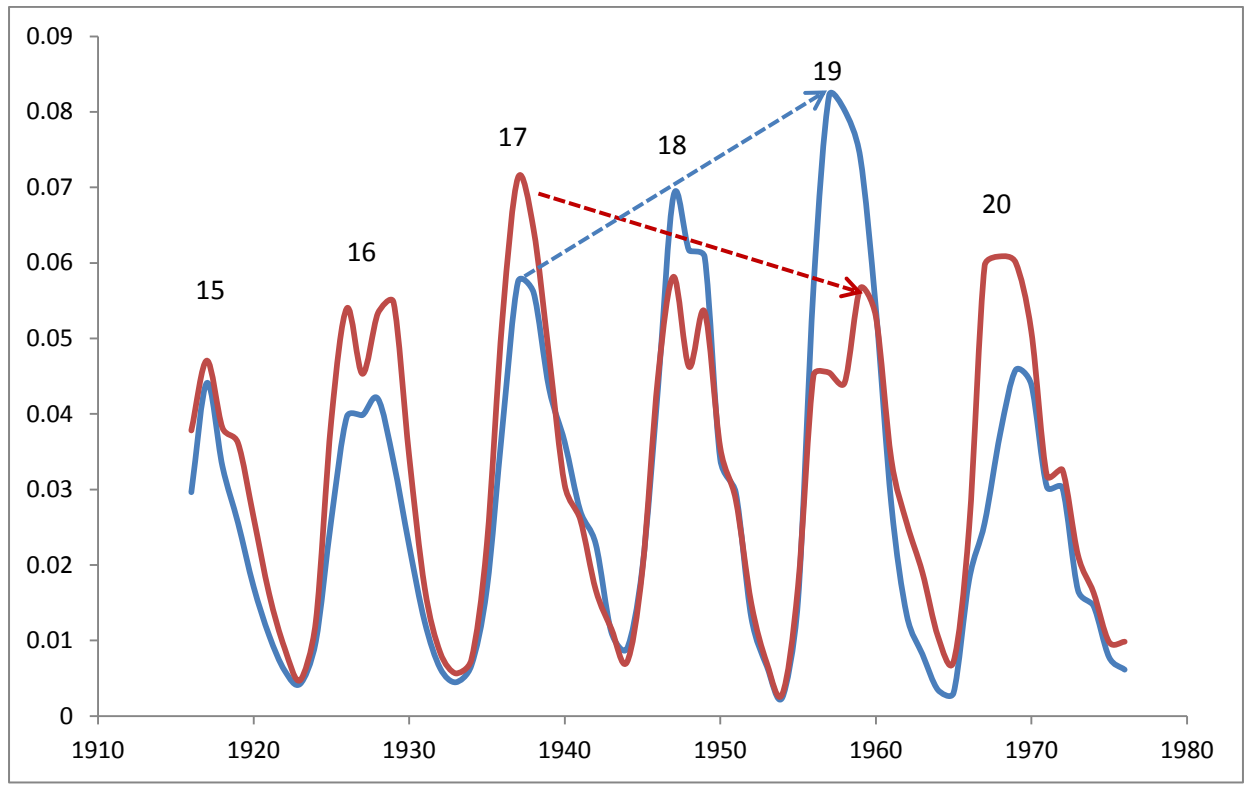

Fig.2 Plots of annual means of $A_{pn}$ (blue) and of $A_f$. Arrows illustrate the opposite trends of the peaks of cycles 17-19 in $A_{pn}$ and $A_f$. Ordinates are in fractions of a solar disk, multiplied by 25 for $A_f$.

This suggests that, at the activity levels near the peaks of cycles 18 and 19, the sun produced progressively more of the larger – diameter flux tubes that form spots, pores and the larger plages, and fewer of the small flux tubes observable as WL faculae near the limb. The similarity between the 10-15% decrease in WL facular area between cycles 17,18 and 19 seen in Figure 2 and the fractional increase in $A_{pn}$ between those cycles supports this interpretation.

These findings encourage confidence in using the 102 year record of WL facular areas compiled by RGO to reconstruct the bright component of TSI variation more accurately than has been possible using the chromospheric indices or the sunspot number.

The first step is to correct the measured TSI variation, $\Delta S$, for the dimming by sunspots. We form annual means of the daily time series of $\Delta S$ compiled by Fröhlich [17] for the 1980 – 1999 period when $\Delta S$ measurements and $A_{pn}$ are both available.

The correction for spot blocking is carried out by adding to $\Delta S$ the time series of $P_s = 0.33\ A_s$, where $A_s$ is the projected spot area in millionths of the solar disk and the factor 0.33 expresses the photometric contrast of a spot. For the period 1980-1999 we use the NOAA/USAF values of $A_s$ multiplied by 1.1 to account for the neglect of spots too small to draw in that data set[18]. For 1916-1976 we use the RGO values of $A_s$, but we multiply the contrast by 0.8 since the many small spots included by RGO have much lower contrast than large spots[18].

We seek the regression of $\Delta S + P_s$ upon $A_f$ but the RGO measurements of $A_f$ stopped in 1976 and reliable measurements of S became available only in 1980. So we must first regress $\Delta S + P_s$ upon $A_{pn}$, which is available through 1999. This yields the relation:

$$\Delta S + P_s = 0.028 A_{pn} - 2 \times 10^{-5}, \qquad (1)$$

with a correlation coefficient of $r = 0.95$.

We then regress the annual means of $A_{pn}$ upon $A_f$ for the period 1916-1999 for which both are available. Here $A_f$ is the projected area of WL faculae as tabulated by RGO.

This regression yields the relation:

$$A_{pn} = 23 A_f - 0.0007 \qquad (2)$$

with a correlation coefficient of $r = 0.84$.

Substituting the relation (2) in (1) we obtain:

$$\Delta S + P_s = 0.64 A_f - 4 \times 10^{-5}. \qquad (3)$$

In making the substitution of (2) into (1) we assume that the relation between $A_f$ and $A_{pn}$ is stationary between the two periods 1916-1976 and 1980-1999. This is a reasonable assumption since the two periods cover a similar range of solar activity, except for a few years around the peak of cycle 19. In any case, neither the shape nor the relative amplitude of the TSI variations calculated from (3) are sensitive to the value of the slope. The magnitude of the variations changes with the slope but this has little importance for our conclusions.

We now use the relation (3) with the time series of $A_f$ to generate the annual means ($\Delta S + P_s$) for 1916-1976. This is the period over which we can compare $\Delta S$ generated from $A_{pn}$ and $A_f$, since both are available. Finally, we subtract $P_s$ from the time series $\Delta S + P_s$, to form $\Delta S$ for 1916-1976.

This time series is shown in Figure 3(a). We see that the TSI amplitudes of the moderately sized cycles 15, 16, 17 and 20 are in roughly the same proportion in $A_f$ as in $A_{pn}$. But the large maxima of TSI generated in the late 1940's and 1950's by the high values of the chromospheric index around the peaks of cycles 18 and 19 are absent in the reconstruction from WL faculae. Instead TSI begins to dip around 1940 to a deep minimum in the late 1950's.

The different behavior is illustrated more clearly in Figure 3(b) where the sunspot cycle variation has been removed with an 11-year running mean. The amplitude of the dip in the annual means is about 0.1% when measured relative to the adjacent 11 – year peaks. It is about 0.06% in the 11 year smoothed time series.

This solar dimming, rather than brightening, at the highest activity levels agrees with stellar photometry which shows that somewhat younger Sun – like stars darken around the peaks of their activity cycles[15]. Over most its activity range the Sun falls into the category of stars that brighten at the peak of their spot cycles18 and 19, the Sun's spatial spectrum of small (bright) and large (dark) magnetic cycles. But our findings indicate that the solar dynamo shifts to sufficiently lower spatial wavenumbers to move it into the category of the stars that darken at their cycle peaks.

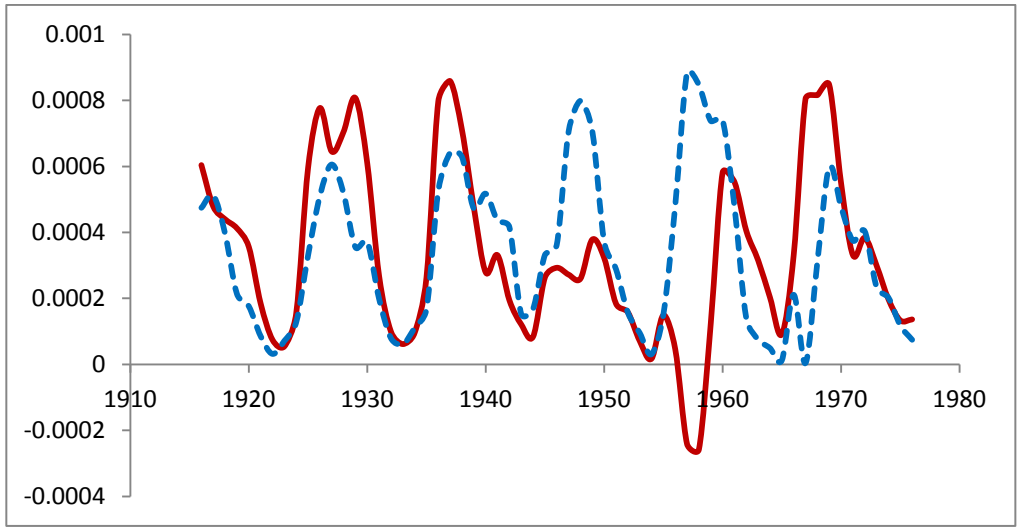

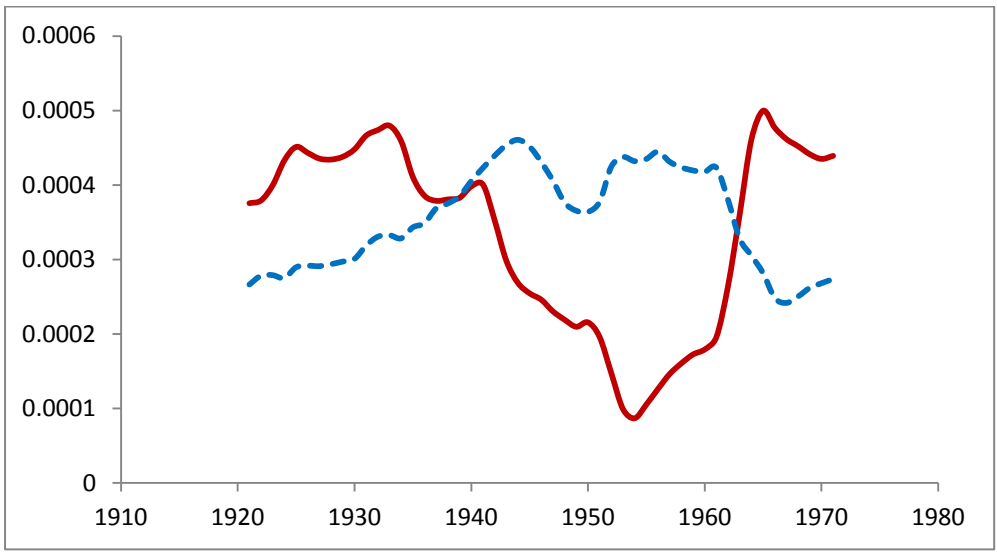

Figure 3: (a) A plot of the annual means of ΔS for 1916 – 1976 reconstructed from $A_f$ (red) and from $A_{pn}$ (blue); (b) same as above smoothed with an 11- year running mean. Ordinates are in fractions of TSI.

The dimming may be somewhat over- estimated because some of the TSI variation is contributed by ultraviolet wavelengths that arise in part from the chromospheres, so they track $A_{pn}$ better than $A_f$. Therefore, the high $A_{pn}$ values during cycles 18 and 19 might increase the minimum TSI in the late 1950's somewhat. The fraction of the TSI that is accounted for by UV variation is controversial, but has been estimated as roughly 10-20% [19].

Suggestions that TSI variation in phase with activity may be caused mainly by UV variation can be discounted because bolometric balloon – borne imaging shows that faculae in and around active regions are bright in evenly weighted integrated light over the wavelength range between about 320 nm and 3 microns[20]. The bolometric contrasts are similar to monochromatic values measured at visible wavelengths that are used in empirical models of TSI variation[21,22]. These can account for TSI variation on the rotational and 11 year time scales. So most of the TSI variation must arise from the visible and IR range.

Whatever the outcome of this controversy, the UV radiation below roughly 320 nm is absorbed by ozone in the stratosphere so it has no direct influence on tropospheric heating. Therefore, the dimming shown in Figures 3 seems to be a good representation of the part of total solar irradiance variation caused by solar activity, as received at ground level.

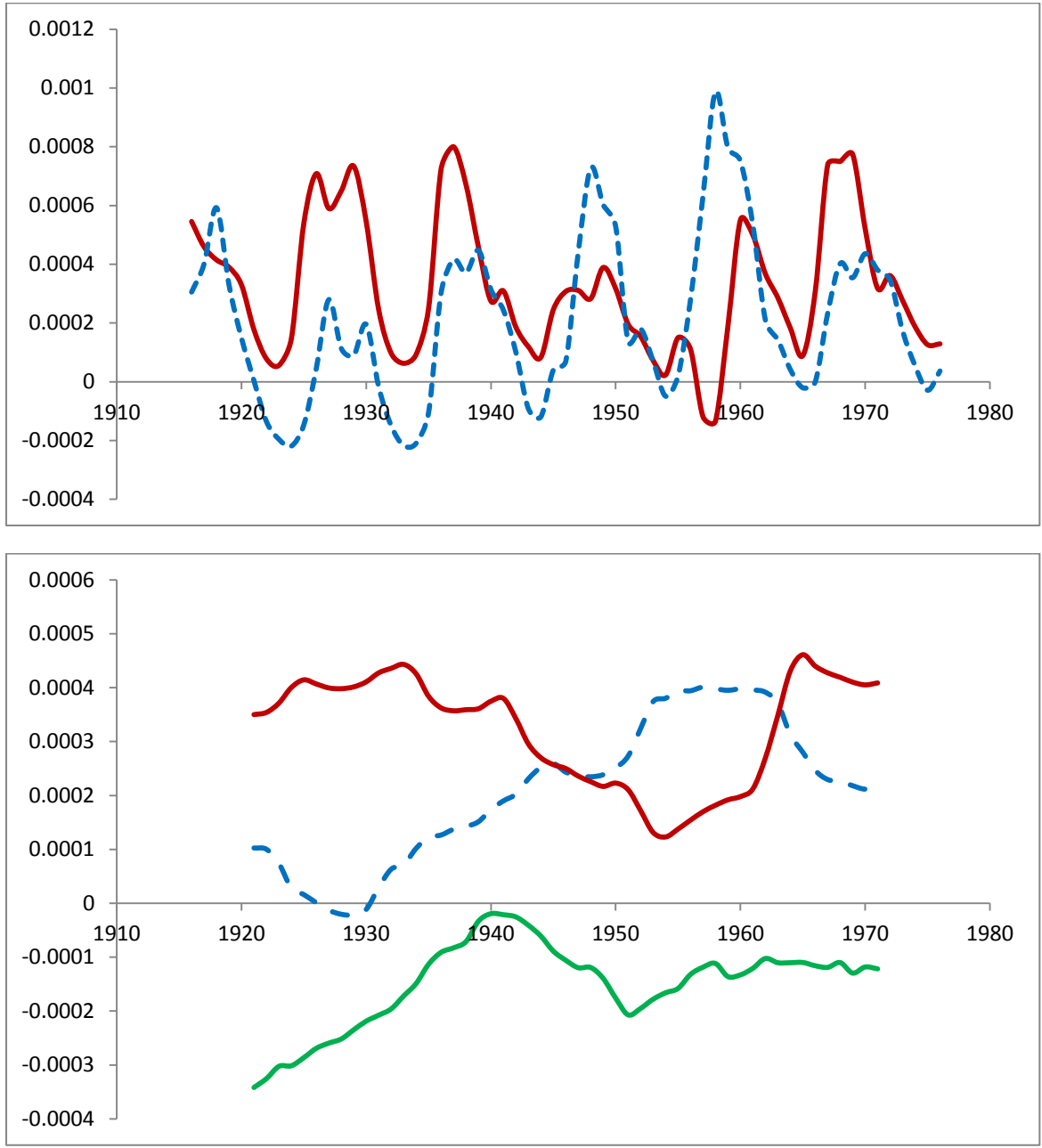

Fig.4(a) Comparison of annual mean $\Delta S$ from $A_f$ (red) and the NRL TSI time series (blue); (b) 11- year smoothed $\Delta S$ from $A_f$ (red), NRL (blue) compared to $T_g$ (green). Ordinates are in fractions of $\Delta S$.

Fig. 4(a) shows a comparison between TSI reconstructed from $A_f$ and the widely used NRL TSI time series. The NRL reconstruction[23] includes a term based on modeling of the turbulent diffusion of photospheric magnetic fields. This additional term accounts for the gradual upward trend in the minima that is not present in the $A_f$ – based curve (nor in the $A_{pn}$ – based curve shown in Fig.3). Except for this upward trend, the NRL series resembles the $A_{pn}$ – based curve in Fig.3; both exhibit peak TSI values around the maxima of cycles 18 and 19, at a time when the $A_f$ – based TSI shows a deep minimum.

Fig. 4(b) illustrates the behavior of the 11- yr smoothed TSI from $A_f$ and from the NRL model, compared to the global- mean surface air temperature record, $T_g$[24]. The global cooling from the early 1940's through 1960's seen in $T_g$, corresponds much better with the $A_f$ – based TSI curve, than with the NRL or $A_{pn}$ – based TSI behavior. This suggests that use of the $A_f$- based curve in climate models is likely to alter the mid 20$^{th}$ century balance between solar, volcanic, aerosol, and greenhouse contributions to global temperature described by the IPCC report[25].

The behavior of $A_f$ is similar to the chromospheric indices (and to the sunspot number) at solar activity levels lower than those experienced in cycles 18 and 19. So our findings do not significantly affect the TSI record since the early 17$^{th}$ century reconstructed from various combinations of R and the chromospheric indices, except during the period of those two high cycles [23, 26,27]. They do, however, imply that an *increase* in future solar activity beyond cycle 19 levels might dim the sun by an amount comparable to, or exceeding the dimming that may have occurred during the 17$^{th}$ century Maunder Minimum of activity.

The similarity in behavior of $A_f$ and $A_{pn}$ except at the highest activity levels limits a direct test of our model against TSI measurements. The Sun's activity level in cycles 21 - 24 since the beginning of space borne TSI measurements has been lower than cycles 18 and 19. So we expect reconstructions of these cycles based on WL faculae and chromospheric indices to be difficult to distinguish. Moreover, the time series of $A_f$ ended with the termination of the RGO measurements in 1976.

Some support is, however, available from the finding that measured *daily* TSI variations increase with sunspot number, R, only up to R~150. Beyond that activity level they begin to decrease [27]. Daily values cover a wider range of variation than annual means, so the saturation and turn -over that causes the TSI decrease studied here is revealed in daily data even at the moderate activity levels in cycles 21 – 24 covered by that study. This turn – over suggests that the TSI amplitude of those cycles was depressed below values that it would have reached if the relation between $A_f$ and $A_s$ were linear. But the amplitudes in spot area of those cycles were too small to cause a pronounced dimming as prolonged as that produced by the larger annual mean spot areas during cycles 18 and 19.

**Acknowledgements**